# Modeling Dynamic Component Interfaces[*]

Franz Huber, Andreas Rausch, Bernhard Rumpe
email: {huberf, rausch, rumpe}@in.tum.de
Technische Universität München
Arcisstr. 21, D-80290 München, Germany

**Abstract**

*In this paper we adopt a component model based on object-oriented systems, introducing the concepts of components and their structure. A component consists of a dynamically changing set of connected objects. Only some of these objects are interface objects, and are thus accessible from the environment. During the component lifetime not only the number of objects, but also that of interface objects, and their connections change. To describe this situation, we introduce Component Interface Diagrams (CIDs) – an adaption of UML diagrams – as a notation to characterize interfaces of components, their structure, and their navigability. We show how CIDs can be used to describe the in-house developed Open Editor Framework (OEF). Finally, we give guidelines that allow to map components described with CIDs directly to several component technologies, like ActiveX, CORBA, or Java Beans.*

## 1: Introduction

Today, on top of object-oriented techniques, an additional layer of software development, based on components is being established. The goals of Componentware [BRS98, Sam97] are very similar to those of object-orientation: reuse of software is to be facilitated and thereby increased, software shall become more reliable and less expensive. One of the goals of the Frisco project [Tec98] that we currently carry out was to develop a framework for graphical and textual editors that was particularly open for the incorporation of new editors, without changing the source code of the framework. In order to achieve this, we decided to use component concepts to structure and encapsulate different entities of the framework, such that they can even be dynamically loaded and unloaded.

Componentware takes a large leap toward reusability, since components aim at a granularity much larger than single objects do. However, today the question what component concepts are, is still under investigation. Several approaches [Sam97] in general agree that components should be based on object-orientation. However, when details are considered some explicit or subtle differences can be found. So, when applying the ideas of components within our object-oriented framework, we first had to clarify our notion of components.

Second, after a clarification what components are, the question was raised how to describe components in an abstract and compact way. While todays methods, like UML [Gro97], as well as their predecessors e.g. OMT [RBP+91], Fusion [CAB+94], or the Booch

---

[*] This paper is joint work of the the project SysLab (supported by the DFG under the Leibniz Program, and by Siemens-Nixdorf), and the project "A1: Methods for Component-Based Software Engineering", a part of "Bayerischer Forschungsverbund Software-Engineering (FORSOFT)", supported by Siemens ZT.

Method [Boo94], offer Class Diagrams that are suited to describe the internal structure of components, Class Diagrams are not quite suited to describe the interface of a component.

Therefore, we introduce a convenient description technique that allows to describe the interfaces of a component, which is called "Component Interface Diagrams". Component Interface Diagrams allow to structure interfaces, define multiplicities, and describe navigation paths between these interfaces that can be used to retrieve new sub-interfaces. Since the capabilities of components that we want to describe are similar to Class Diagrams, we adapted the latter for our needs.

Our Component Interface Diagrams emerged during the development of the framework FRISCO and greatly helped to define a good architecture. This technique was subsequently generalized and proved useful within other developments. The quality of the FRISCO framework was considerably improved by the notion of components that we introduced.

In the remainder of this section, we briefly introduce the FRISCO OEF framework, which will serve as example application, and discuss the properties of components on which we base our notation. Component Interface Diagrams are introduced and applied to the FRISCO framework in Section 2. In Section 3 we discuss a mapping of our component concept to common object technologies, such as ActiveX [Cha96], CORBA [OHE96], and especially Java Beans [Mic97].

### 1.1: A brief introduction into FRISCO OEF

FRISCO is a document-oriented software engineering tool prototype. It is based on a subset of UML notations [Gro97] but incorporates precisely defined refinement and transformation rules. FRISCO provides a variety of editors combining graphical and textual parts as well as tables within a single document. An example of a FRISCO editor is given in Figure 1.

To achieve flexibility, we developed the OEF (**O**pen **E**ditor **F**ramework) as an open approach of nesting document parts into one compound document. The developed framework provides a standardized set of protocols for embedding documents. To structure these protocols, our notion of component interfaces is used.

For each document element, a specific kind of editor, called *PartHandler*, exists. Each *PartHandler* component consists of a possibly large set of internal objects implementing its functionality. A subset of these objects provides the protocol interface necessary for embedding it into the enclosing document frame. The interface objects hide the internal object structure of a *PartHandler*. They are the only way of communication with the environment. This framework, which has deliberate similarities to OPENDOC [App96], is implemented in Java, and the *PartHandlers* are realized as Java Beans.

### 1.2: Properties of our component-based model

The concept of components is built on top of object-oriented concepts. This allows to use all advantages of object-orientation and build the component layer in such a way that programming in the large is even more feasible. When looking at the implementation of a component, we find the usual object structure. However, if a set of objects that commonly performs a task is grouped together, a new kind of entity with new characteristics emerges, which needs a new and appropriate kind of description.

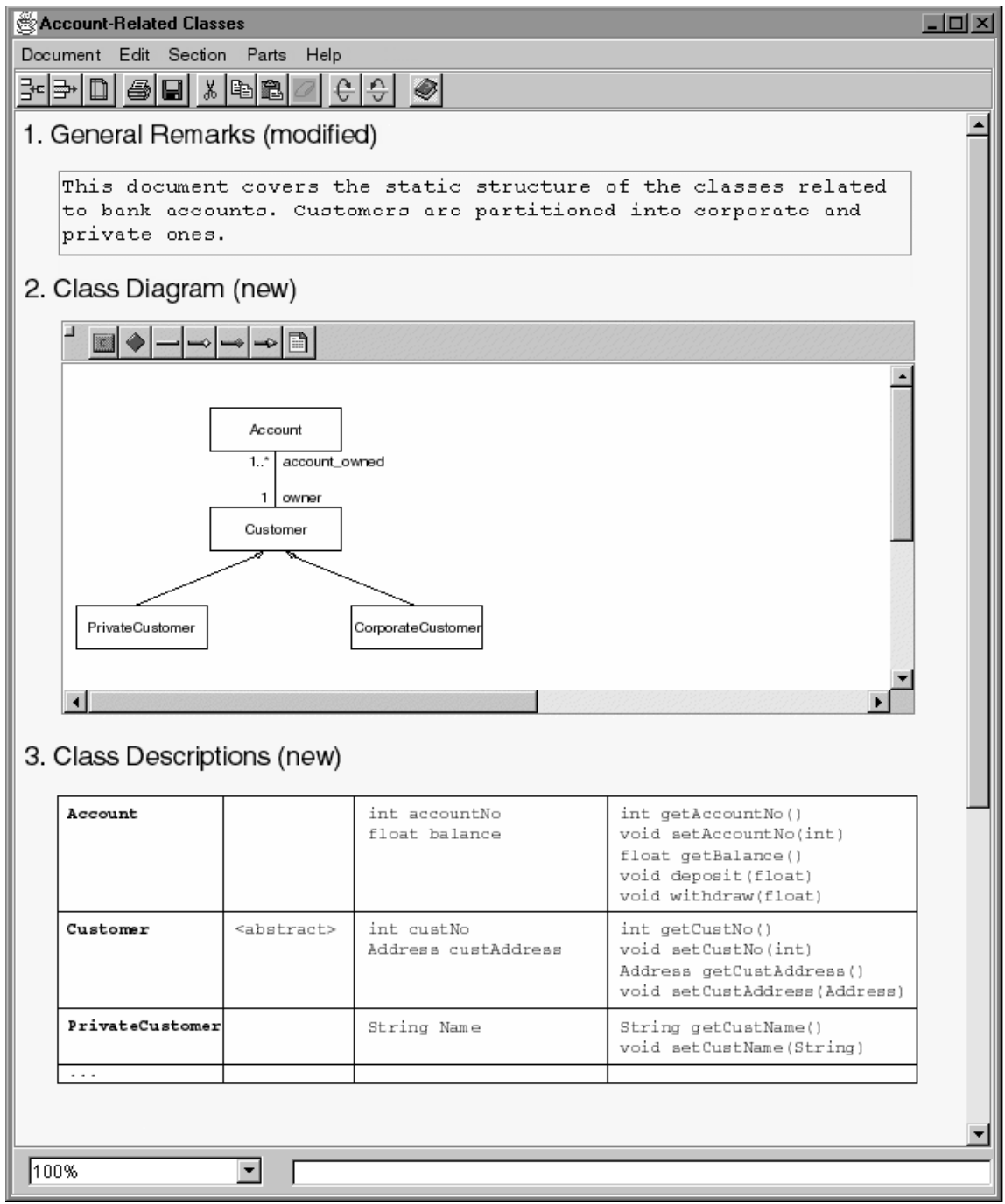

**Figure 1. A Sample Screenshot of a Compound Document Editor in** FRISCO

We do not enforce every entity of the system to be considered a component, but allow independent objects to live between components. Thus developers are free to choose what they want to be a component. Components may interact directly, but may also be glued together using independent objects.

The component concept fits into the type system of the underlying language, such as in Java [GJS96]. As components are intended to be reused across language boundaries, there could be a mapping of the component infrastructure into several type systems as, e.g., found in CORBA.

Components exhibit a characteristics similar to objects. Their instances can be dynamically created, they have a clearly defined interface, and they have a well-structured state. Beyond objects, they exhibit some additional features. A component has hierarchically structured interfaces, hierarchically structured states, and state and interface structure may change dynamically.

To achieve this, we assume a component to consist of a dynamically changing set of objects, that are either internal to the component or are part of its interface.

A so-called *principal object* controls the components. The lifecycle of the component instance is exactly the lifecycle its principal object. Other components and objects can initially access a component via the principal object. From the principal object they can receive references to other interfaces of the component. This way, a complex interface structure to access the component can be obtained.

Once a reference of an internal object has been given to the environment, this object is no longer internal, but belongs to the interface of the component. Thus, the interface of the component is dynamically changing.

Our experiences show that, in many cases, it is not necessary to use concepts of object migration between components. Since component-based systems usually have a rather static structure, it is sufficient to allow objects that have been internal to a component to "appear" on the interface, thus allowing to access them from outside. Please note that we regard physical distribution and migration completely independent of the logical structure. Thus a part of a component may migrate between systems, but still be a part of the (now physically distributed) component. In general, it is not necessary for components to be tightly connected, e.g., allowing to realize object factories [GHJV94].

Objects that are created within a component belong to this component during their lifetime. We assume that objects are not explicitly destroyed but garbage collected which allows us to disregard dangling references and related problems.

## 2: Describing components

Using the concept of components during software development, it is important to have appropriate modeling techniques at hand, that directly allow to deal with components. The newly developed standard UML [Gro97] provides a rich set of techniques for describing different views of objects. Especially useful for describing components are the following notations:

**Interaction Diagrams** describe interactions either between objects in a component, or between components.

**State Diagrams** are a descendant of StateCharts [Har88] and characterize the behavior of single objects within a component, but also of an abstraction of the entire component's behavior.

**Interface and Class Declarations** describe the methods and attributes, together with their types and access rights.

**Class Diagrams** are used to describe the possible structures of a system or a component.

**Object Diagrams** define the static part of the internal structure of a component.

Our experiences show that a larger subset of the objects within a component has the same lifecycle as the principal object and does not change its linkage. Thus, the internal structure of a component is rather static and can be described by an Object Diagram.

However, when regarding components UML does not directly provide sufficient techniques to describe the interface structure of a component. As the interface structure of a component consists of a dynamically changing set of objects, it is increasingly important to have an appropriate notation to give an abstract and compact overview of this interface structure. Beyond the given UML notations, we propose in Section 2.2 an adapted version of Class Diagrams – *Component Interface Diagrams* – that allows us to cope with the extended capabilities of component interfaces.

### 2.1: FRISCO OEF interfaces

In FRISCO OEF several kinds of components are used. We now introduce and briefly describe a subset of the interfaces that *PartHandler* components provide.

**BasicPartHandler** is the principal interface that every *PartHandler* must provide. It covers rudimentary content and embedding functionality and allows to access additional interfaces of a *PartHandler*. To allow the enclosing document frame to access part information relevant for embedding, a number of methods are available to obtain information about content and size. Please note that this interface does not provide services for editing documents, since it is desirable that certain document parts should be displayed read-only.

**Edit** interfaces can be obtained by invoking the *getEdit* method. This interface is provided only if the part is editable. It basically provides the services to externalize (save) its content and to activate and deactivate editing capabilities.

**Menu** interfaces allow access to the *PartHandler*'s menus. Two menus are allowed (one attached to the global menu bar, and a contextual menu).

**Undo** allows a *PartHandler* to participate in the OEF Undo/Redo mechanism. After an *ActionListener* registers at the component, it receives a *UndoableAction* each time a change occurs.

**Connection** allows to access the interconnections between *PartHandlers* in the compound document, e.g., to propagate changes in order to ensure consistency between parts.

### 2.2: Purpose of Component Interface Diagrams

At the beginning of the lifetime of a component, the principal object (in FRISCO an instance of *BasicPartHandler*) is the only object that is accessible from the environment. Thus the interface of the component is initially given by the principal object. Over time, this may change. More objects may be created inside the component, and a reference to them may be given to the environment, leading to a dynamic extension of the component interface. This provides an important component property: being able to provide additional interfaces during runtime if required. The purpose of a *Component Interface Diagram* (CID) is to give clients a concise knowledge of the possible set of interfaces they may use and how access to these interfaces can be gained.

Due to the requirement of strong typing for components, interfaces may be created during runtime, but their type must be known initially. A CID gives information about the visible interfaces, their inheritance relations, and navigation paths between these interfaces. Furthermore, methods and multiplicities of these interfaces are shown. CIDs are adapted from UML *Class Diagrams*. Figure 2 shows an CID for the *PartHandler* component.

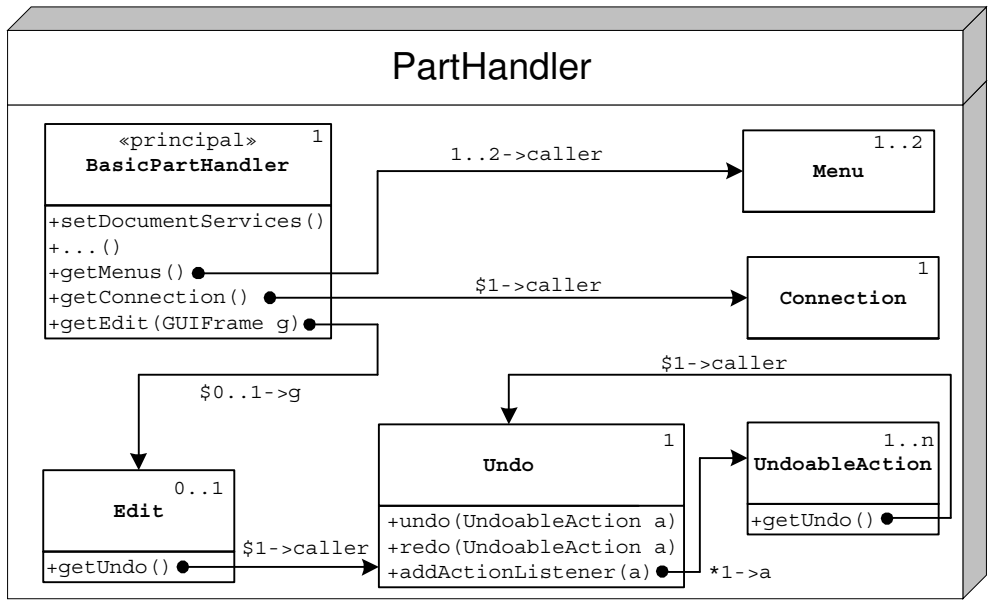

**Figure 2. A** FRISCO **Component Interface Diagram**

Let us for now disregard the arrows and their labels. Besides denoting the kind of components a CID belongs to (here *PartHandler*), a CID contains externally visible classes, their inheritance relations, visible methods, and, in addition, multiplicities of possible instances.

The *PartHandler* in Figure 2 offers six externally visible interfaces, among them the principal interface marked with the appropriate stereotype.

Each class can be given a multiplicity that determines the maximum allowed set of interfaces during runtime. Each interface will usually be implemented by an object, and a component may provide multiple objects of the same class at its component interface. Like many other components *PartHandler* has several single instance-only interfaces (e.g. the *Edit* interface), but also unconstrained ones, like *UndoableAction*.

Public methods together with their type may be provided in the CID just like they can be defined within UML's Class Diagrams. The + in front of each method just indicates that it is public as it is in UML. The method list serves two purposes. A method often either is used to provide navigation facilities from one interface to another, or realizes functionality provided by the component. Although the latter is the more important, we now focus on the former.

Navigation paths are introduced as a concept to indicate the possible paths where to navigate from one interface to another. A navigation path is denoted by an arrow from a method of one interface to another interface. The *PartHandler* in Figure 2 shows, which navigation paths between interfaces are existing. It tells us, e.g., that from the *Edit* interface, the *Undo* interface can be obtained.

Navigation between interfaces is done by calling the method, usually resulting in a reference to a new interface (see Section 2.3 for a detailed discussion). Please note that these navigation paths are not the same as associations. Although an association is a good candi-date to be the component's internal way to implement efficient support for such navigation, it is left open to the components internal details how to support navigation. Another way to implement navigation is to create a new object with the appropriate interface each time such a navigation access is required.

So far the CIDs give a first flavor of the interfaces of a component, but their expressiveness is limited. We therefore add a transition labeling to describe how new interfaces can be obtained, whether we iteratively receive the same interface, or a new one for each request. For example, calling *getMenus* on the principal interface returns one or two *Menu* interfaces to the caller (`1..2->caller`).

The multiplicities on navigation arrows indicate how often the use of this method leads to a new interface. In general they do not tell us what happens, if the method is called too often.

However, if iterative calls result in the same interface for all callers, this is indicated by "$". To indicate the creation of a new interface "*" is used instead (see method *addActionListener*).

The communication between a component is often not limited to a call from the environment and a return from the component. Instead, when called, a component can itself make "call backs" to objects of the environment. By these call backs, additional objects can become externally known, without the initial caller of the components method being involved. Such an example is given by the call of *getEdit* that does not return an interface to the caller but passes the method's parameter along to another call returning this interface to the object referenced by the parameter (`$0..1->g`). Please note, that such a "call back" in general need not take place immediately, but can be delayed (e.g., done by another thread). Furthermore, repeated call backs are allowed. For instance, the *Undo* interface allows to register *UndoActionListeners* (method *addActionListener*) that will receive a reference to an *UndoableAction* each time an undoable change occurs.

CIDs specify which references to its objects a component can give to the environment. A careful flow analysis, as done for other purposes already in Java compilers, could prove correctness of the component implementation.

There are basic objects, such as Java Strings, that are publicly available (see Section 1.2). It is useful to exclude such basic classes from the component concept, but to let them float through component borders freely, regardless, where they have been created. However, such exclusion has to be done carefully, being aware of implicit communication via shared objects which could lead to a behavior that is not derivable by observation of component interfaces.

Given the technique of Component Interface Diagrams and the already mentioned notations of UML, we can define different views of components. With CIDs, we can define the *Black-Box View* of components. Class Diagrams are useful to specify the internal structure of a component, the so-called *Glass-Box View*. With object diagrams we can specify run-time behavior of components as an object structure snapshot. Figure 3 illustrates the relationship between these different diagrams: With class diagrams one can show the implementation of CIDs (see Section 3). Object diagrams can be used to show run-time behavior of class diagrams.

Hence, the integration of a CID within standard object-notations like UML can be given by a mapping of the CID into an embedding class diagram, where all component interfaces map to classes, the inheritance relation and the multiplicities are preserved, and the navigation relation is mapped to method calls accordingly.

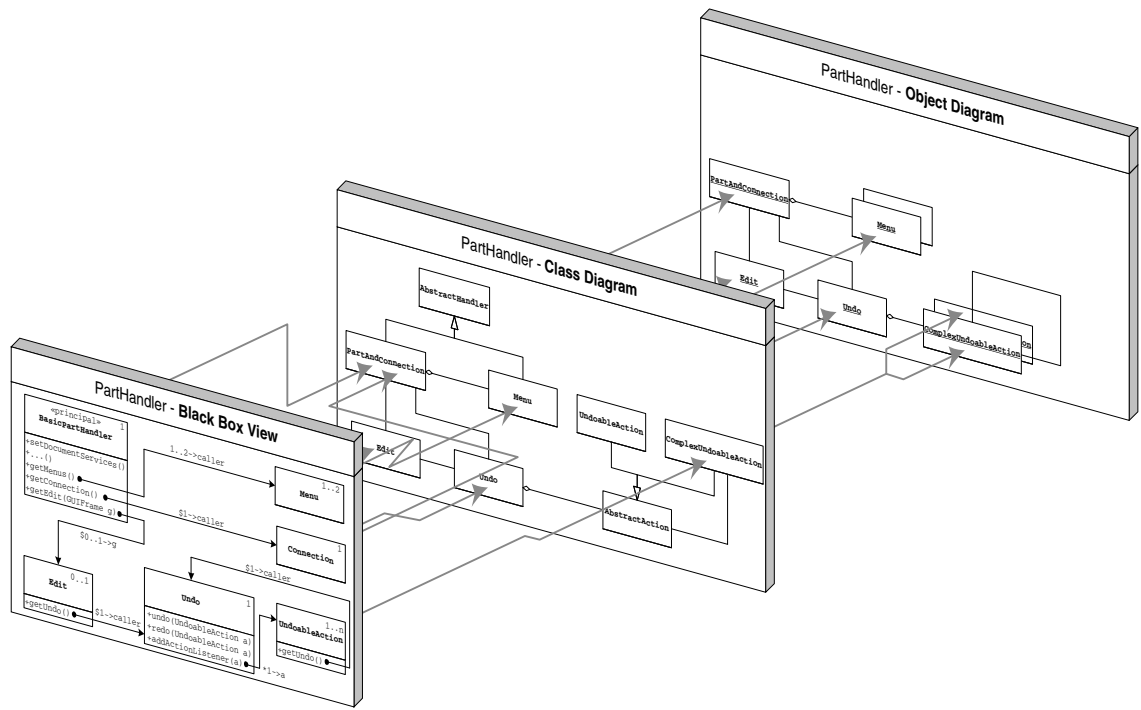

**Figure 3. From Component Interface Diagrams over Class Diagrams up to Object Diagrams**

The most important capability of components is the possibility to provide a complex, well-structured set of individual and standard interfaces. Therefore, a classification of interfaces is a point of interest following two main goals:

- Separation of concerns for the component developer ending up with a more modular implementation than one monolithic interface could provide.
- Clearly structured individual and standard interfaces to give component users a more natural way of understanding the different purposes of the entire component.

The designer of a CID should structure the interfaces with respect to appropriate methodical guidelines. This could be expressed in UML stereotypes for standard interfaces. For example, special interfaces for storage, printing, the undo/redo-mechanism, security, configuration, online help, testing and debugging are often useful. These standard interfaces are especially needed for component-based systems supporting plug-in of components, like, for instance, editors with exchangeable spell checkers.

**2.3: Guidelines to map components to objects**

Based on our experiences, we suggest the following guidelines for a mapping. In general, there are three kinds of possibilities to implement navigation between interfaces.

We have focused on the preferable *method call*. But it is also possible to use public readable *attributes* for interface access if they are available, or a dynamic *cast* of a given interface into another interface. The latter is, e.g., possible in Java, where failed casts can be caught by an exception.

Component interface types are mapped either into Java classes or Java interfaces. The former has the disadvantage that classes are not abstract and thus can be instantiated from the environment, the latter cannot be used if attributes are publicly available in the interface. As we prefer methods for navigation, we suggest to use Java interfaces to implement CID interfaces.

When the desired multiplicity of an interface is 1 or a link has modifier $, then the interface needs to be stored after creation within the component to be repeatedly exported. Its creation can either be done when the component is created, or in a lazy manner, when the first request is served. Anyhow, these interfaces should be implemented following the singleton pattern [GHJV94].

If multiplicity of a navigation, or of an interface is restricted and repetition is not wanted, at least the number of already created interfaces needs to be stored. A proper reaction for too many requests is necessary: either returning `nil` or throwing an exception. The standard for too many requests is the latter one, the former one should be used to cope with optional interfaces.

The creation of a component goes along with the creation of its principal object. For that purpose, the creator must know the actual class of the principal object. It therefore helps to use the same names for the component and the principal class. In our example we did not follow this principle in order to simplify discussion: The component *PartHandler* and the basic interface resp. class *PartHandler* (here called *BasicPartHandler*) are something quite different. If clients want to instantiate components in a flexible way, a global name service or an object factory [GHJV94] should be implemented.

A navigation path will often be realized using an association. However, such an association relates objects within the component and therefore is part of the implementation and not of the interface of the component. Although associations are good candidates for navigation path implementation, this is not enforced. Another way to implement navigations is to use variables that are global within the component when, e.g., multiplicity is set to 1. Yet another way to implement navigations is possible if a new interface is created and given to the environment with each method invocation. These new interfaces need not be stored within the component, but can themselves contain references to other component parts.

Similar to aggregation of objects, we conceptually allow the hierarchical composition of components. However, our experiences show, that in practice, components will not be deeply nested. The composition of components is done by creating and using a component within another one.

## 3: Mapping the component model to component infrastructures

Today, three main component infrastructures are in practical use: Microsoft's *ActiveX*, based on OLE and DCOM [Cha96], several *CORBA* implementations [OHE96], and SUN's *Java Beans* [Mic97]. It is difficult to estimate at this time which technology will dominate in the future. Consequently, there should be a mapping of CIDs in all three technologies available.

As all three technologies support a composition concept and provide an interface definition language – MS-IDL, IDL, and Java Interfaces – , a CASE tool supporting CIDs or similar description techniques could generate interface definitions for each technology.

Hence, a mapping from our component based model to these technologies is basically possible.

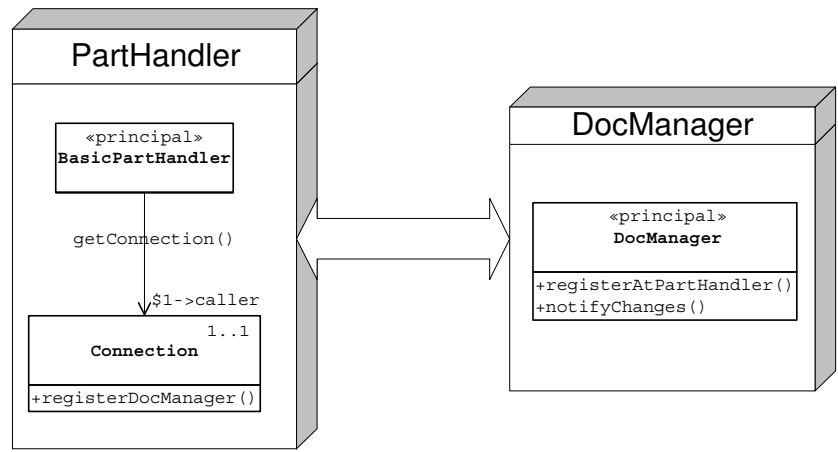

**Figure 4. Interacting** OEF **Components**

Representatively, we discuss a Java Bean-based implementation of the component-based system shown in Figure 4. This system presents an abstraction of two FRISCO components: The *PartHandler* (see Section 2.1, Figure 2) and a new component, the *DocManager*. The purpose of the *DocManager* is to observe its *PartHandlers* and propagate changes to related *PartHandlers*. If the method *registerAtPartHandler* is called the *DocManager* receives a pointer to the *Connection* interface (*getConnection*) and registers itself (*registerDocManager*). Afterwards, if a user edits any diagram, the corresponding editor component (*PartHandler*) notifies the *DocManager*, which then ensures that all other affected *PartHandlers* are informed of the change, eventually disallowing it, if it leads to inconsistent documents.

### 3.1: Implementing Component Interface Diagrams with Java Beans

According to its creators from JavaSoft "A Java Bean is a reusable software component that can be manipulated visually in a builder tool" [Mic97, JT98]. This covers a wide range of different possibilities. The scope of functionality reaches from simple GUI parts, like buttons, up to full-featured database access adaptors.

In technical terms, a Bean is a Java object. The specific characteristics of Beans are:

**A Public Interface** offers *Properties*, *Methods*, and *Events* for clients to access the Bean.
**Introspection** allows a builder tool to explore the Bean's interfaces and present it to programmers. For that purpose, the *Java Reflection Technique* is used.
**Customization** allows developers to change the properties of Beans during design-time.
**Persistence** is used to store the Bean's state permanently and restore it later.

Beans can support additional features, such as, e.g., security, drag & drop, or remote invocation. To support several of these features, Beans have to obey some conventions.

As Beans are just Java objects, Beans can implement several Java interfaces. This fits directly into our component concept, as we also allow several interfaces for each component and inheritance between interfaces. Beans also support single inheritance, which is not yet used for components in our model.

Beans are packaged in so-called JAR files that include, among code and other resources, optionally serialized Bean instances. Components in our component model can be connected via links. As the standard Java name service is a crude circumvention to establish links between Bean instances in different JAR files, it is necessary to define an own name service, or to use the new *Java Naming and Directory Interface* [Jav98], or even to use a Bean-conformant infrastructure supporting a global name service, like, e.g., IBM's ComponentBroker [IBM98].

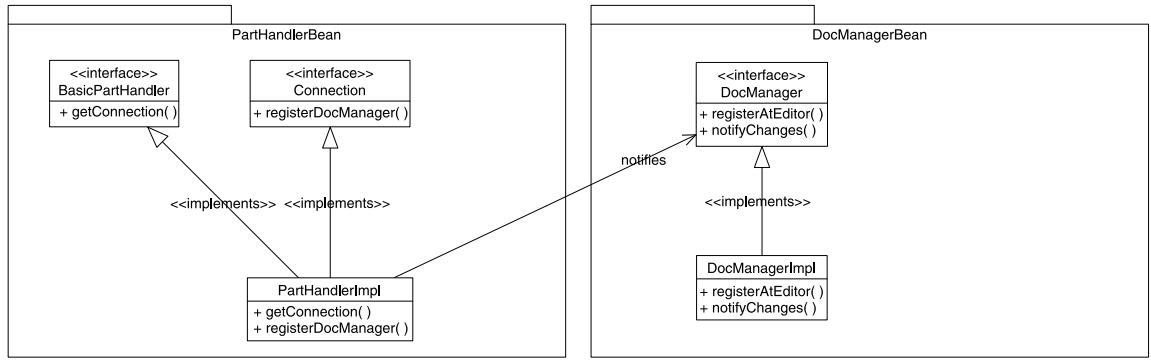

**Figure 5. Implementing CIDs with Java Beans**

Each CID interface is mapped into a Java interface, for each CID component a Java Bean is realized, where the Java Bean has to implement the corresponding Java interface. Figure 5 illustrates an implementation of the example given in Figure 4 by using a conventional UML Class and Package Diagram. There are three Java interfaces, one for each CID interface (*BasicPartHandler*, *Connection*, and *DocManager*). In the presented solution there are only two Java classes implementing these three interfaces. One could also implement the interfaces with more classes or provide additional attributes, methods, or classes, like, for instance, the class *PartHandlerImpl*, which has an association to the interface *DocManager*. The Java classes and interfaces are packaged into JAR files, one for each CID component.

The resulting JAR files resemble the implementation of former components, thus the *PartHandler* component is implemented as a package named *PartHandlerBean*.

### 3.2: Implementing Component Interface Diagrams with ActiveX or CORBA

Implementing CIDs with other component infrastructures, like ActiveX or CORBA, is similar to what has been presented in the previous section with Java Beans. In ActiveX, each component can provide several interfaces, which maps directly into our model. But ActiveX does not support the concept of subtyping, so subtyping should not be used in CIDs if the target is ActiveX. Since ActiveX provides only a simple naming service, called *Monikers*, we suggest to implement an own naming service or use standardized implementations, as, e.g., provided in CORBA, to realize links between ActiveX components.

Using CORBA as target means having interfaces that allow multiple inheritance, as proposed in our model. But a CORBA object cannot implement more than one interface. Instead, CORBA offers a module concept where interfaces can be grouped together into a specific namespace, given by the surrounding module. Hence, in CORBA, CID components are thus reduced to simple namespaces. But CORBA provides a global name service. Links between CORBA objects as needed in our component model can be implemented in a straightforward fashion.

## 4: Related work

Our work is inspired from work found in the area of Architecture Description Languages (ADL) [HHK+96], the OPEN Modeling Language (OML) [FHSGPJ97], Catalysis [DW95], and UML [Gro97]. In the field of ADL descriptions of components and their interactions are found. OML as well as UML also provide description techniques for components and interfaces and their collaboration. Finally, Catalysis offers description techniques for components supporting several interfaces and their relations.

In contrast to CIDs, where one specifies a single component, its interfaces, and the corresponding navigation paths, all of these description techniques describe a set of components and their interactions. With CIDs one can specify a component and its interfaces without describing the concrete context, namely other components using the described component. This is especially useful in Componentware since components are intended to be reused in different environments.

To sum up, most component-related description techniques describe components and their interactions in a specific context. With CIDs it is possible to describe components without a specific context, concentrating on their interfaces and navigation paths.

## 5: Conclusion

The proposed concept of components was defined as a result of designing and implementing the FRISCO framework for document editing. Although UML provides several description techniques to describe different views of object-oriented systems, including component implementations, a need for the description of component interfaces arises. To remedy this problem, Component Interface Diagrams have been introduced. They are essentially an adaption of UML Class Diagrams for purposes of describing structured and dynamically changing interfaces and their navigation paths.

The high quality of FRISCO shows the suitability of the component concept and the defined notation. Although several extensions are imaginable, e.g., allowing object migration or defining a notion of inheritance on components (not only their interfaces), we expect the given notion of components and the defined concept of Component Interface Diagrams to be sufficient for a large class of applications.

We consider it to be more important that language and tool support allow to conveniently define component types and automatically translate them into object-oriented implementations. This would considerably boost component technology.